\documentclass[osajnl,twocolumn]{revtex4} 
\usepackage{graphicx}
\begin{document}

\title{Low-frequency measurements of electro-optic coefficients of $\mathbf{Na}$:$\mathbf{KTiOPO_4}$ and $\mathbf{RbTiOAsO_4}$}
\author{Daruo Xie and Olivier Pfister}
\address{University of Virginia, Department of Physics, \\382 McCormick Road, 
Charlottesville, VA 22904-4714, USA}
\email{opfister@virginia.edu}

\begin{abstract}
We report the measurement of the $r_{23}$ and $r_{33}$ electro-optic coefficients of the nonlinear optical crystals $\rm Na:KTiOPO_4$ (Na:KTP) and $\rm RbTiOAsO_4$ (RTA). We observed a marked decrease of the electro-optic effect in the Na:KTP crystals for frequencies below 100 Hz. This effect is absent in RTA. 
\end{abstract}

\maketitle 

\section{Introduction}
In recent work, our group demonstrated an ultrastable, continuous-wave, nondegenerate optical parametric oscillator (OPO), based on an $x$-cut Na:KTP crystal. The frequency difference of the OPO signal and idler beams was controlled to sub-hertz accuracy and continuously tunable between 0 Hz and hundreds of MHz.\cite{SPIE} This was achieved by electronically locking the optical phase difference of the two OPO beams. In this lock, the correction signal was sent to electrodes across the Na:KTP crystal, generating a field along its $z$ axis. The electro-optic (EO) effect was thus used to tune the OPO frequencies to exact degeneracy, thereby generating quantum mechanically indistinguishable, photon-number correlated optical beams, which were used in the first demonstration of macroscopic Hong-Ou-Mandel quantum interference.\cite{PRL} In the course of this experiment, we observed an unexpected decrease of the EO tuning with the frequency of the phase-lock error signal. This was tested by applying an EO sinusoidal modulation to the OPO crystal and monitoring the resulting FM spectrum on the signal-idler beat note, which revealed a dramatic diminution of the modulation depth as the modulation frequency decreased below a few kHz. This observation in our Na:KTP OPO crystal provided the incentive for the present investigation of the EO effect at low frequencies in flux-grown Na:KTP and RTA. Note that the observed levels of quantum noise reduction (squeezing) in our Na:KTP crystals were quite similar to what is expected for KTP,\cite{PRL} hence we do not expect that the anomalous behavior of the electro-optic effect could be a consequence of absorption in Na:KTP, since the latter would decrease squeezing as well.

We used two methods to measure the dependence of the electro-optic coefficients on the frequency of the applied field. In the first method, the crystal was placed inside an optical resonator and the EO effect was easily observed by measuring the cavity resonance shift with an oscilloscope. The EO coefficients $r_{23}$ and $r_{33}$ of the crystal were then individually measured by polarizing the input light field along the $y$ and $z$ axes of the crystal, respectively. Although sensitive, due to the amplification of the EO effect from the multiple passes of the resonant field through crystal, this measurement of the resonance shift was nevertheless fairly noisy. We therefore used an ellipsometry setup in order to confirm the initial results. This ellipsometry measurement was more precise than the first method, even though it only enabled access to the composite EO coefficient $n_z^3r_{33}-n_y^3r_{23}$. The two methods produced consistent results for Na:KTP.

\section{Experiments and results}

\subsection{Cavity resonance modulation measurements in Na:KTP}

\begin{figure}[htb]
\begin{center}
\includegraphics[width=3.25in]{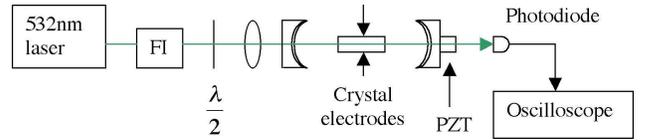}
\end{center}
\caption{Experimental setup for the cavity resonance modulation EO measurement. FI: Faraday isolator. PZT: piezoelectric transducer. The voltage $V_o\sin\omega t$, where  $V_o= 365$ V, was applied along the $z$-axis of the crystal, giving a maximum field $E_z$ of $1.2 \times 10^5$ V/m. The cavity length was scanned at 40Hz by the PZT.}
\label{cav}
\end{figure}

The experimental setup is shown in Fig.\ref{cav}. 
The optical cavity was composed of two concave mirrors of 150 cm curvature radius and 99\% reflectivity. The incident light at $\lambda = 532$ nm was emitted by a frequency-doubled Nd:YAG monolithic laser (Lightwave Electronics, model 142). We used an $x$-cut, 5 mm $\times$ 3mm $\times$ 3mm flux-grown Na:KTP crystal fabricated by Coherent Crystal Associates and AR coated at 1064 and 532 nm. The crystal was placed at the center of the cavity. An AC voltage was applied to the top and bottom $x$-$y$ surfaces of the crystal, onto which were attached Copper foil electrodes. Hence, an electric field $E_z$ was established along the $z$ crystal axis. The cavity's optical length is
\begin{equation}
L_k = l_o + n_k l,
\end{equation}
where $k=y,z$, depending on the polarization of the beam, $l_o$ is the cavity length in air, $l=5$ mm is the length of the Na:KTP crystal ($l_o+l=10$ cm), and $n_k$ is the index of refraction. 
The cavity free spectral range is 
\begin{equation}
\Delta_k = \frac c {2L_k}
\end{equation}
and the cavity resonance frequency $\nu_k = m\Delta_k$, where $m$ is an integer. The EO index change is given by\cite{yariv}
\begin{eqnarray}\label{eoy}
\delta n_y& = & \frac{n_y^3}2\ r_{23}\ E_z\\
\delta n_z& = & \frac{n_z^3}2\ r_{33}\ E_z \label{eoz}
\end{eqnarray}
Since the EO effect is small, one can write the cavity resonance change as
\begin{equation}
\delta\nu_k \simeq -\frac{2ml}c\ \Delta_k^2\ \delta n_k.
\end{equation}
Using Eqs.\ (\ref{eoy},\ref{eoz}) in the previous equation, one obtains
\begin{eqnarray}
|r_{23}| &=& \frac{c}{n_y^3 l\Delta_y\nu_y}\ \frac{\delta\nu_y}{E_z}\\
|r_{33}| &=& \frac{c}{n_z^3 l\Delta_z\nu_z}\ \frac{\delta\nu_z}{E_z}.
\end{eqnarray}
The modulation amplitudes $\delta\nu_k$ and the free spectral ranges $\Delta_k$ are measured on a digital oscilloscope, taking great care not to choose commensurable frequencies for the field modulation and the PZT scan. The free spectral ranges for different light polarizations are taken equal to first-order approximation. 

\begin{figure}[htb]
\begin{center}
\includegraphics[width=3.25in]{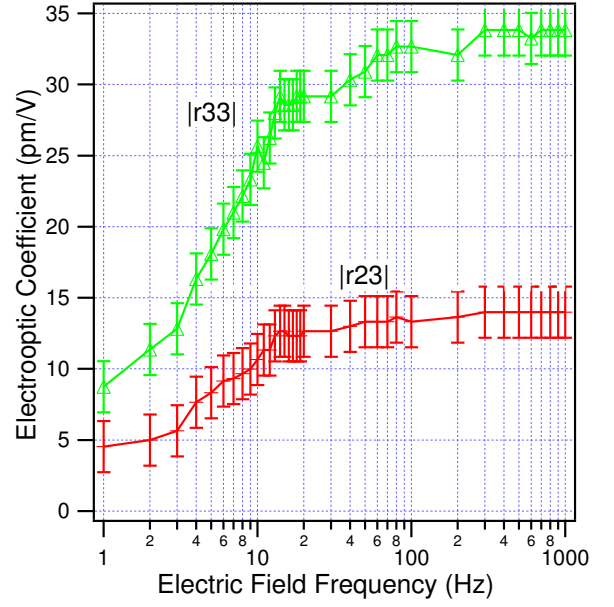}
\end{center}
\caption{Experimental determination of the magnitudes of the Na:KTP EO coefficients $|r_{23}|, |r_{33}|$, versus EO modulation frequency.}
\label{cavres}
\end{figure}

Figure \ref{cavres} displays the measurement results.
One clearly sees that the EO coefficients of Na:KTP drop sharply when the frequency of the applied electric field is below 20Hz. At higher frequencies, the measured values for $r_{23}$ and $r_{33}$ in Fig.~\ref{cavres} are in agreement with the respective values of 15.7 pm/V and 36.3 pm/V, previously measured at 633 nm for KTP.\cite{ref} 

Although giving access to the individual EO coefficients, this method is fairly noisy and inconvenient (it could be improved by use of lock-in detection, for example). In order to be able to conveniently study and compare different crystals, we subsequently adopted a different approach. 

\subsection{Ellipsometry measurements in Na:KTP and RTA}

\begin{figure}[htb]
\begin{center}
\includegraphics[width=3.25in]{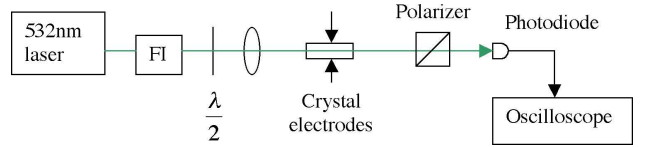}
\end{center}
\caption{Experimental setup for the ellipsometry measurement. FI: Faraday isolator. The voltage $V_o\sin\omega t$, where  $V_o= 380$ V, was applied along the $z$-axis of the crystal.}
\label{ell}
\end{figure}

\begin{figure}[htb]
\begin{center}
\includegraphics[width=3.25in]{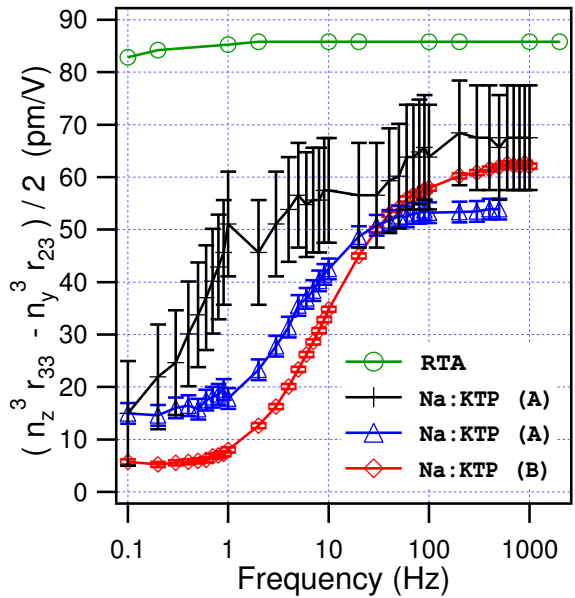}
\end{center}
\caption{Experimental results of the ellipsometry measurements. The Na:KTP (A) trace with cross markers and large error bars corresponds to the determination of $|\delta n_y-\delta n_z|/E_z$ from the results of Fig.~\ref{cavres} (cavity resonance modulation).}
\label{ellres}
\end{figure}

The experimental setup is shown in Fig.\ref{ell}. 
The polarization of the incident light is now linear at a $\rm 45^o$ angle from $y$ and $z$. The relative phase shift between the two polarizations yields an elliptically polarized beam. Crossed polarizers thus allow us to measure the changes of the degree of ellipticity caused by the EO modulation in the crystal. 
For an input field 
\begin{equation}
\tilde E(0) = E_0 (\hat y + \hat z) e^{-i\omega t},
\end{equation}
the output field is 
\begin{equation}
\tilde E(l) = E_0 \left(\hat y + e^{i\theta}\hat z\right) e^{-i\omega t+i\Theta},
\end{equation}
where $\theta = 2\pi(n_z- n_y) l /\lambda$ and $\Theta = 2\pi n_yl/\lambda$. In the diagonal basis $\hat d_{\pm}=(\hat y\pm\hat z)/\sqrt 2$, this gives
\begin{equation}
\tilde E(l) = E_0 \left[\left(1+e^{i\theta}\right) \hat d_+ + \left(1-e^{i\theta}\right)\hat d_-\right] e^{-i\omega t+i\Theta}.
\end{equation}
The beam intensity of the polarization transmitted by the output polarizer ($\hat d_-$) has therefore the dependence $|1-e^{i\theta(t)}|^2=2-2\cos\theta(t)$, which allows us to determine the EO modulation $\delta\theta\propto \delta n_z-\delta n_y=(n_z^3 r_{33} - n_y^3 r_{23})E_z/2$ from Eqs.\ (\ref{eoy},\ref{eoz}). We tuned the average $\theta$ to $\pi/2$.  Figure \ref{ellres} displays the measurement results for two different Na:KTP crystals and one RTA crystal. 
The two Na:KTP crystals differed in that one (crystal A) was a new crystal, whereas the other (crystal B) had been previously used in a continuous-wave OPO. The crystal sizes were 5 mm $\times$ 3 mm $\times$ 3mm for the Na:KTP crystals and 20 mm $\times$ 3 mm $\times$ 1 mm for the RTA crystal. All crystals were flux-grown by Coherent Crystal Associates.

In both Na:KTP crystals, the EO effect is seen to disappear at low frequency. In addition, results of Fig. \ref{cavres} have been plotted on Fig.\ref{ellres} for comparison: it is clear that the two methods give consistent results. RTA, remarkably, has a quasi constant EO effect at all frequencies. We also found that the measured value of $|\delta n_y-\delta n_z|/E_z$ for RTA is very close to the one determined from the RTA EO coefficients measured in Ref.~\cite{ref}.

\section{Conclusion}

We have observed a decrease of the EO efficiency below 100 Hz modulation frequency for Na:KTP crystals. RTA crystals do not show this effect. We hypothesize that an explanation for this decrease in the electro-optic effect might be related to internal field decrease due to ion migration in the crystal, from superionic conductivity induced by site hopping of K or Na ions in crystals with imperfect stoichiometry.\cite{jiang} A broadband measurement of electrical conductivity, beyond the capabilities of our current home-made autobalancing bridge, may be useful to gain more insight in the process.\cite{archer,jiang} These results are of importance for the realization of ultrastable OPO's for quantum optics applications, including continuous-variable entanglers for quantum information\cite{rd} and ultraprecise measurements.\cite{ol} We find that RTA (or periodically poled RTA) crystals are better suited than Na:KTP for these systems, with similar nonlinear coefficients and no EO abnormalities.

\section*{Acknowledgements}

This work was supported by NSF grants No. PHY-0240532 and No. EIA-0323623.

\end{document}